\documentclass[showpacs,prl,amsmath,amssymb,superscriptaddress,nobibnotes, twocolumn,preprintnumbers]{revtex4-1}

\usepackage{graphicx}
\usepackage{endnotes}
\usepackage{bm}
\usepackage{epsfig}
\usepackage{amsmath}

\newcommand{\SrIr}{Sr$_{3}$Ir$_4$Sn$_{13}$}
\newcommand{\SrRh}{Sr$_{3}$Rh$_4$Sn$_{13}$}

\newcommand{\CaSrRhx}{(Ca$_{x}$Sr$_{1-x}$)$_3$Rh$_4$Sn$_{13}$}
\newcommand{\CaSrIrx}{(Ca$_{x}$Sr$_{1-x}$)$_3$Ir$_4$Sn$_{13}$}
\newcommand{\CaRh}{Ca$_{3}$Rh$_4$Sn$_{13}$}
\newcommand{\CaIr}{Ca$_{3}$Ir$_4$Sn$_{13}$}

\newcommand{\Tstar}{$T^*$}

\begin{document}

\title{Evidence of a structural quantum critical point in (Ca$_{x}$Sr$_{1-x}$)$_3$Rh$_4$Sn$_{13}$ \\from a lattice dynamics study}
\author{Y.~W.~Cheung}
\author{Y.~J.~Hu}
\affiliation{Department of Physics, The Chinese University of Hong Kong, Shatin, New Territories, Hong Kong, China}

\author{M.~Imai}
\author{Y.~Tanioku}
\author{H.~Kanagawa}
\author{J.~Murakawa}
\author{K.~Moriyama}
\affiliation{Department of Chemistry, Graduate School of Science, Kyoto University, Kyoto 606-8502, Japan}

\author{W.~Zhang}
\author{K.~T.~Lai}
\affiliation{Department of Physics, The Chinese University of Hong Kong, Shatin, New Territories, Hong Kong, China}

\author{K.~Yoshimura}
\affiliation{Department of Chemistry, Graduate School of Science, Kyoto University, Kyoto 606-8502, Japan}

\author{F.~M.~Grosche}
\affiliation{Cavendish Laboratory, University of Cambridge, J. J. Thomson Avenue, Cambridge CB3 0HE, United Kingdom}

\author{K.~Kaneko}
\affiliation{Materials Sciences Research Center, Japan Atomic Energy Agency, Tokai, Naka, Ibaraki 319-1195, Japan}

\author{S.~Tsutsui}
\affiliation{Japan Synchrotron Radiation Research Institute (JASRI), SPring-8, Sayo, Hyogo 679-5198, Japan}

\author{Swee~K.~Goh}
\email{skgoh@phy.cuhk.edu.hk}
\affiliation{Department of Physics, The Chinese University of Hong Kong, Shatin, New Territories, Hong Kong, China}
\affiliation{Shenzhen Research Institute, The Chinese University of Hong Kong, Shatin, New Territories, Hong Kong, China}
\date{29 June 2018}

\begin{abstract}

Approaching a quantum critical point (QCP) has been an effective route to stabilize superconductivity. While the role of magnetic QCPs has been extensively discussed, similar exploration of a structural QCP is scarce due to the lack of suitable systems with a continuous structural transition that can be conveniently tuned to 0~K. Using inelastic X-ray scattering, we examine the phonon spectrum of the nonmagnetic quasi-skutterudite \CaSrRhx, which represents a precious system to explore the interplay between structural instabilities and superconductivity by tuning the Ca concentration $x$. We unambiguously detect the softening of phonon modes around the {\bf M} point on cooling towards the structural transition. Intriguingly, at $x=0.85$, the soft mode energy squared at the {\bf M} point extrapolates to zero at $(-5.7 \pm 7.7)$~K, providing the first compelling microscopic evidence of a structural QCP in \CaSrRhx. 
The enhanced phonon density-of-states at low energy provides the essential ingredient for realizing strong-coupling superconductivity near the structural QCP.

\end{abstract}



\maketitle

The emergence of superconductivity near a quantum critical point has inspired intensive research on the role of quantum fluctuations on stabilizing new phases. While the temperature-tuning parameter phase diagrams constructed for a wide range of superconductors share qualitative similarity \cite{Mathur1998,Slooten2009,Gegenwart2008,Paglione2010,Shibauchi2014,Gruner2017}, the pairing mechanism highly depends on the type of the quantum critical point (QCP). 
The magnetic QCP, for example, is generally believed to be related to the surrounding unconventional superconducting state, where the pairing may not be mediated by phonons \cite{Scalapino2012,Monthoux2007}.  To find out if quantum fluctuations from other types of QCP can enhance superconductivity, the identification of QCPs associated with a non-magnetic transition -- and the search of superconductivity in the vicinity -- is an important topic in correlated electron research.

Approaching structural instabilities has been another promising route to enhance superconductivity. This concept has been applied to materials such as A15 compounds \cite{Chu1973,Chu1974,Testardi1975}, cuprates \cite{Holder1992,Birgeneau1987,Wakimoto2004}, perovskites \cite{Kang2011}, transition metal dichalcogenides \cite{Yang2012,Pyon2012,Fang2013,Kamitani2016,Kudo2016,Heil2017}, Ni- and Fe-based superconductors \cite{Cruz2008,Yoshizawa2012,Niedziela2011,Kudo2012,Hirai2012}. 
However, contrary to the magnetic counterpart, clear examples of a QCP resulting solely from a tunable structural phase transition are rare, thereby hampering a systematic study of structural quantum criticality and its influence on superconductivity. Recently, the existence of the structural QCP (SQCP) has been suggested in LaCu$_{6-x}$Au$_x$ \cite{Poudel2016} and ScF$_3$ \cite{Handunkanda2015}, although superconductivity has not been observed near the SQCP in the former compound, and the latter compound is an insulator.

The nonmagnetic quasi-skutterudite superconductors \CaSrRhx\ and \CaSrIrx\ have recently been identified as promising systems for realizing a SQCP. The end compound \SrRh, with a space group of $Pm\bar{3}n$ at room temperature \cite{Kase2011}, undergoes a second-order structural transition to $I\bar{4}3d$ at $T^*=138$~K and a superconducting transition at $T_\mathrm{c}=4.2$~K \cite{Goh2015,Yu2015,Lue2016,Cheung2016,Cheung2017}. With an increasing Ca fraction $x$, which introduces chemical pressure, $T^*$ decreases linearly and extrapolates to 0~K at $x\approx0.9$ \cite{Goh2015,Yu2015,Hu2017}, while $T_\mathrm{c}$ takes a dome-shaped dependence with a maximum value of 8.9~K. At $x=0.9$, the coupling strength is found to be dramatically enhanced ($2\Delta_{\rm sc}/k_\mathrm{B}T_\mathrm{c}=6.36$), and a distinct linear-in-$T$ resistivity was detected from $T_\mathrm{c}$ to $\sim$40~K, accompanied by a 33\% reduction of the Debye temperature compared with $\Theta_\mathrm{D}=244$~K at $x=0$ \cite{Goh2015,Yu2015}. Therefore, the phase diagram of \CaSrRhx\ is highly suggestive of a SQCP, around which a broad superconducting dome is located. \CaSrIrx\ shows a similar phase diagram, except that the full suppression of $T^*$ requires a physical pressure of $\approx$18~kbar on \CaIr\ \cite{Klintberg2012,Biswas2015}. 

Nevertheless, microscopic evidence for the existence of a SQCP has thus far been missing. The feature associated with the structural transition in a range of experimental probes is weakened when approaching the putative SQCP. In \CaSrRhx\ with $x = 0.75$, which is still quite far away from the SQCP, the structural transition can no longer be identified in specific heat \cite{Yu2015}, and only a weak feature can be discerned in electrical resistivity \cite{Goh2015}. Indeed, for structural transition in metallic system, it seems to be a common issue that it is difficult to identify the transition when tuning towards the SQCP. Examples include Lu(Pt$_{1-x}$Pd$_x$)$_2$In \cite{Gruner2017} and 2$H$-NbSe$_2$ \cite{Feng2012} which have structural transitions driven by charge-density-wave instability. This would raise questions on the nature of the lattice dynamics close to the SQCP. 

In this Rapid Communication, we investigate the lattice dynamics in \CaSrRhx\ with four calcium concentrations using inelastic X-ray scattering (IXS). We discover that, at $x=0.85$, the soft mode energy squared at the {\bf M} point extrapolates to zero at $(-5.7 \pm 7.7)$~K. Thus, the collapse of an energy scale is detected, providing the first microscopic evidence for the existence of a SQCP in this series. Our data offer a natural explanation for the observation of low $\Theta_\mathrm{D}$  and the strong-coupling superconductivity in the vicinity of $x=0.9$.
\begin{figure}[!t]\centering
      \resizebox{9cm}{!}{
              \includegraphics{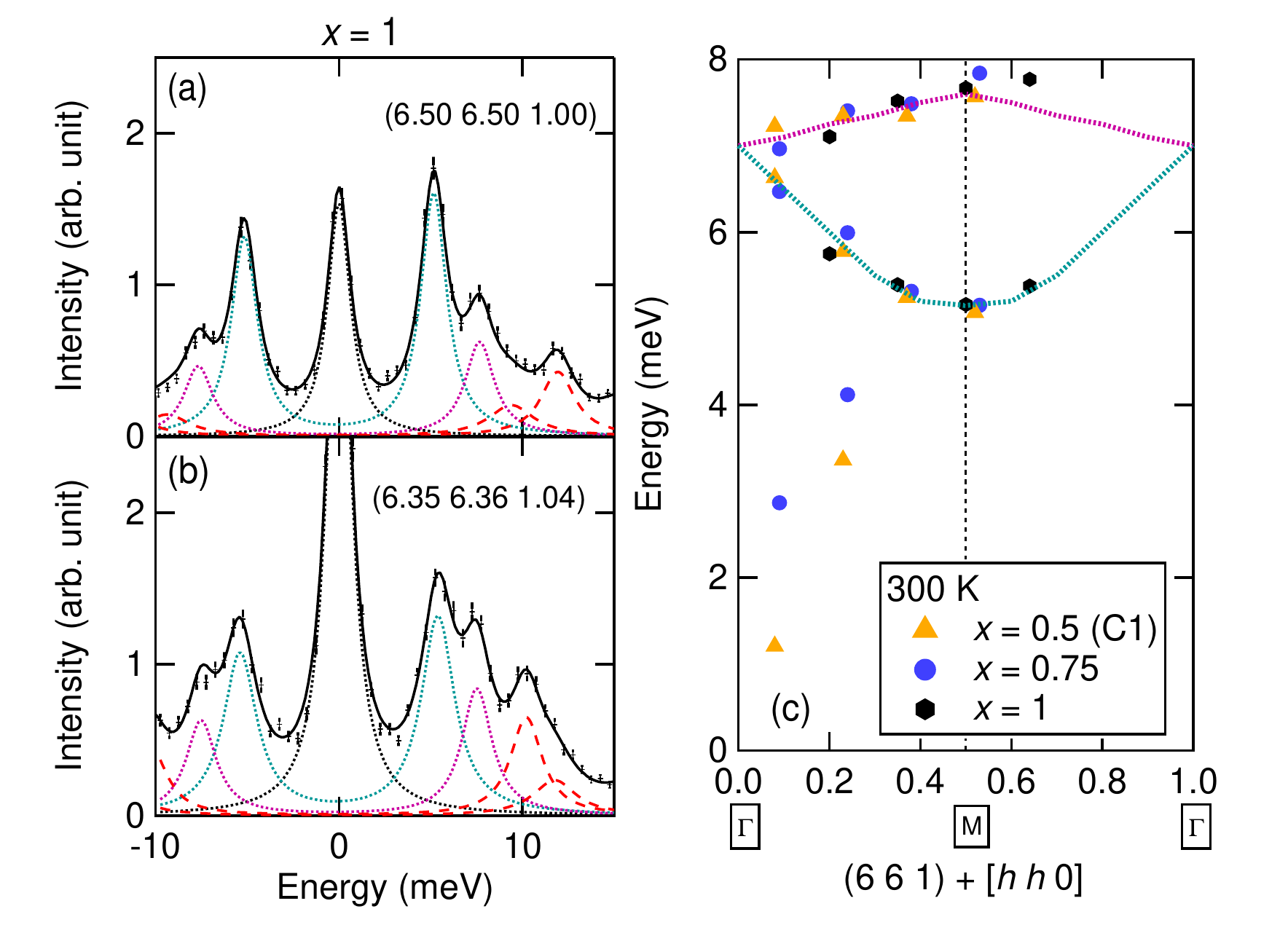}}                				
              \caption{\label{fig1} 
              Room temperature IXS data showing intensity against energy transfer for $x=1$ at (a) $\mathbf{Q}=$(6.50 6.50 1.00) and (b) (6.35 6.36 1.04) . Data are shown by crosses with vertical error bars. The black solid lines are the fits described in the Supplemental Material \cite{SUPP}. (c) Dispersion relations showing the phonon modes with energies lower than 8~meV for various calcium contents. For $x=1$, these are the modes displayed as dotted lines in (a) and (b). The dotted lines in (c) are guides for eyes. 
              }
\end{figure}


Single crystals of (Ca$_x$Sr$_{1-x}$)$_3$Rh$_4$Sn$_{13}$ were obtained by the Sn flux method as described in Ref. \cite{Yang2010}. 
IXS was performed at BL35XU of SPring-8 in Japan. Further details of the experimental setup and the analysis of the IXS spectra are provided in the Supplemental Material \cite{SUPP}.



\begin{figure}[!t]\centering
       \resizebox{9cm}{!}{
              \includegraphics{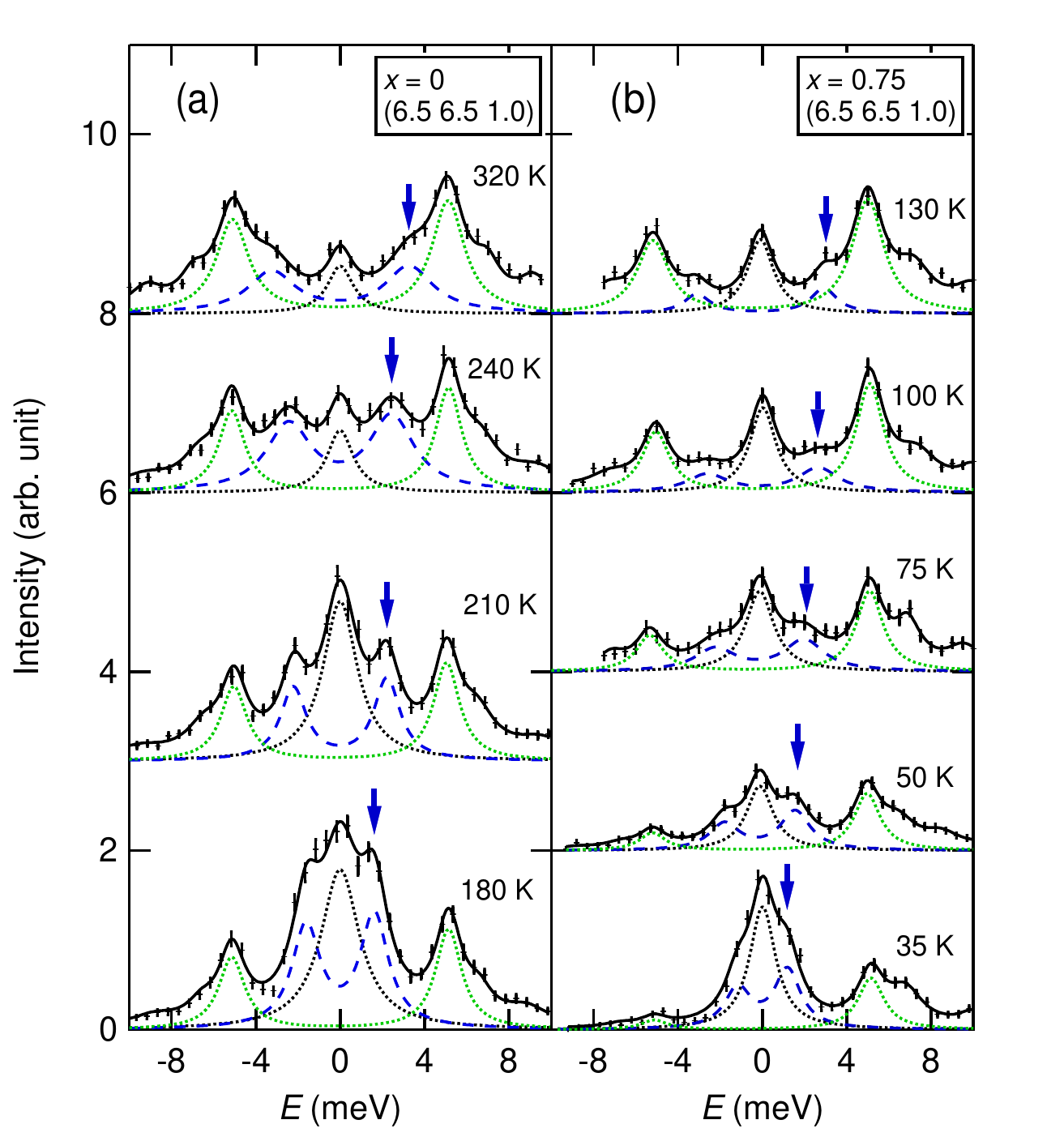}}              
              \caption{\label{fig2} 
              Constant-$\mathbf{Q}$ scans at the {\bf M} point with temperatures above the structural transition temperature in (a) $x=0$ and (b) $x=0.75$. The blue dashed lines represent the soft phonon mode responsible for the structural transition. For clarity, only the two peaks with the lowest energies are shown.
              }
\end{figure}
To understand the evolution of the phonon modes, we first examine \CaRh\ ($x=1$) at 300~K. In \CaRh, the room temperature $Pm\bar{3}n$ cubic structure remains stable down to the lowest temperature \cite{Goh2015, Yu2015}. Hence, \CaRh\ at 300~K is far away from structural instability and the phonon structure serves as an important benchmark. Figs.~\ref{fig1}~(a),(b) show representative constant-$\mathbf{Q}$ scans at $\mathbf{Q}$=(6.50 6.50 1.00) and (6.35 6.36 1.04) for \CaRh\ at 300~K. As shown in both figures, we successfully resolve the elastic peak and four independent inelastic peaks below 15~meV. The inclusion of the peaks coming from the anti-Stokes process improves the reliability of the fit. The inelastic peaks thus resolved are plotted as individual Lorentzian peaks, with the center corresponding to phonon mode energies \cite{SUPP}. 

The complex unit cell, with 40 atoms in each structural unit, gives a large number of phonon branches. We find that the peaks contributed by phonons with energy higher than $\sim$8~meV highly overlap in energy space within our resolution. However, the main purpose here is to track the evolution of low-energy excitations, and hence a complete resolution of high energy modes is beyond the scope of the present study. By plotting the energies of the two peaks with lowest energies along the ${\bf \Gamma}$-{\bf M} direction, as shown in Fig.~\ref{fig1}(c), we identify them as two optical branches. After overlaying the dispersions at 300~K for $x=0.75$ and $x=0.5$ (C1) on the same graph, it is immediately clear that they all share the same optical branches below 8~meV, indicating that these branches are not sensitive to chemical pressure. In the constant-$\mathbf{Q}$ scans near ${\bf \Gamma}$, $x=0.75$ and $x=0.5$ (C1) show extra but weak intensities at low energy transfer, which can be attributed to acoustic branches. 


Next, we investigate the lattice dynamics on approaching structural instabilities. Previous density functional theory calculations for \SrRh\ and \SrIr\ \cite{Goh2015,Tompsett2014} suggest that the phonon instability is located at the {\bf M} point. Subsequently, softening of the phonon mode at {\bf M} was observed by inelastic neutron scattering in \CaIr\ \cite{Mazzone2015}. In light of these results, we conduct a series of IXS measurements on \CaSrRhx\ with four calcium contents to probe the phonon structure at {\bf M}. Fig.~\ref{fig2} shows the resultant spectra in $x=0$ and $0.75$. Similar data for $x=0.5$ (C2) and $x=0.85$ can be found in the Supplemental Material \cite{SUPP}. Due to the presence of the elastic peak, the peak contributed by the soft mode may be masked near zero energy. With this in mind, we performed measurements covering a wide temperature range. From the decomposition of the total intensity, we can recognize a peak, plotted using green dotted lines, at $\sim$5~meV in all spectra. Therefore, this peak is not sensitive to temperature, despite the fact that the temperature has been varied by a large amount. In fact, this is the same peak that already exists far away from structural instabilities, as discussed above.

Close inspection of the spectra in Fig.~\ref{fig2} reveals the existence of another low-energy excitation, represented by the peak in blue dashed line. In $x=0.75$, for instance, this mode has an energy of $3.0$~meV at 130~K, as indicated by the vertical arrow. With cooling, the mode energy decreases rapidly and reaches 1.2~meV at 35~K. This is in stark contrast to the peak in green dotted line, which exhibits no temperature dependence. For $x=0.75$, it has been established via electrical resistivity that $T^*\approx28$~K \cite{Goh2015}. Therefore, the softening of this mode on cooling confirms its relevance to structural transition.

\begin{figure}[!t]\centering
       \resizebox{9cm}{!}{
              \includegraphics{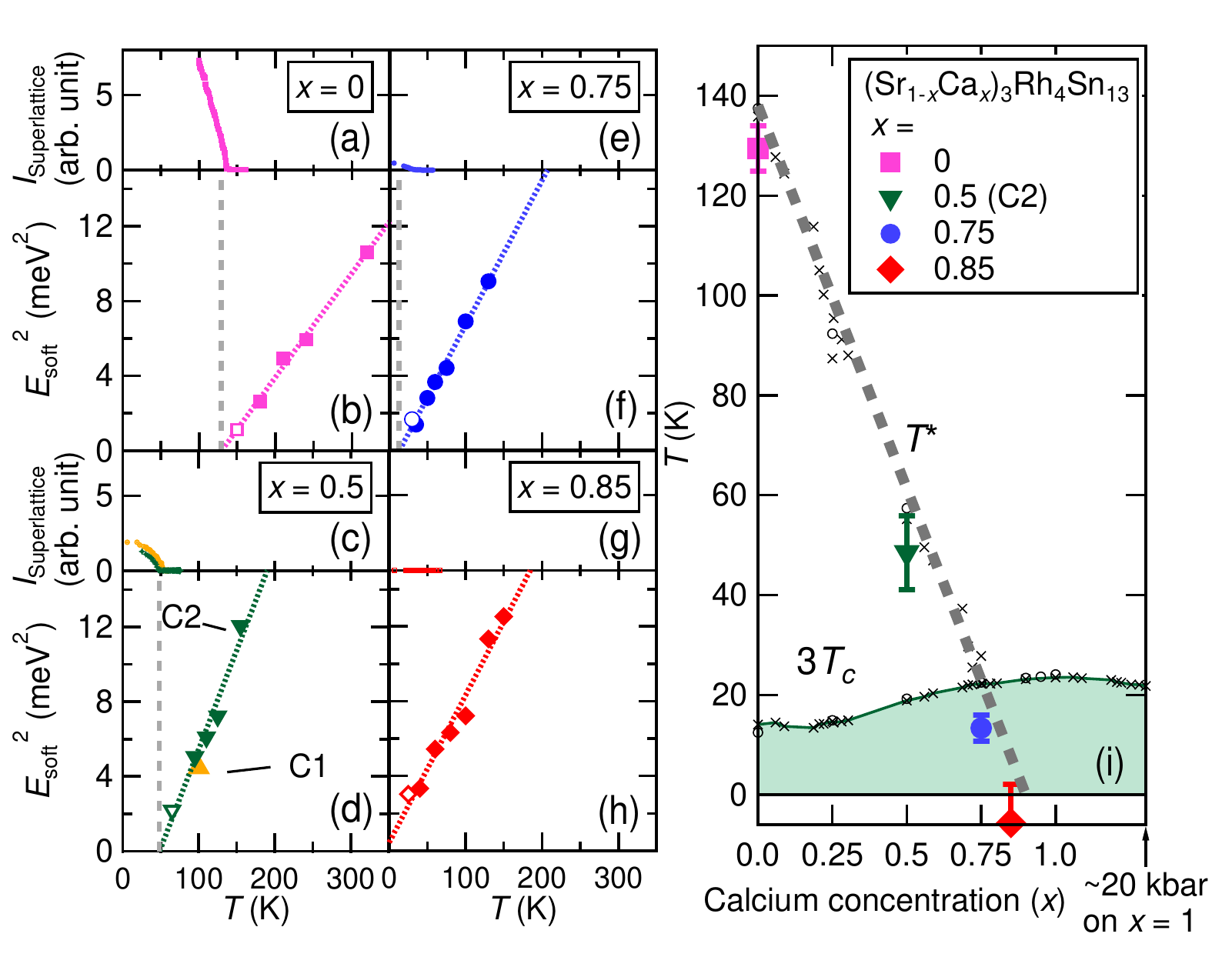}  }  
              \caption{\label{fig3} 
              Intensity of the superlattice reflection and energy squared of the soft mode against temperature in (a),(b) $x=0$; (c),(d) $x=0.5$; (e),(f) $x=0.75$ and (g),(h)$x=0.85$. Two $x=0.5$ crystals (C1 and C2) were measured, showing a small difference, which can be attributed to a slight variation in the actual calcium content. For (b),(d),(f),(h), closed symbols are extracted by fitting the IXS spectra with Equation (S1), open symbols are determined by fitting the dispersion relations with Equation (1). The dotted lines are the linear fit of the data, and the vertical dashed lines mark the $T^*$. (i) Phase diagram of (Ca$_{x}$Sr$_{1-x}$)$_3$Rh$_4$Sn$_{13}$. The polygons represent $T^*$ obtained from IXS data, defined as the temperature where the energy squared of the soft mode extrapolated to 0~K. Small circles and crosses are data from specific heat and electrical resistivity in Ref.~\cite{Yu2015} and Ref.~\cite{Goh2015}, respectively.}
\end{figure}

The energy squared of this soft mode ($E_{\rm soft}^2$) in $x=0.75$ is plotted against temperature in Fig.~\ref{fig3}(f), along with similar data obtained in $x=0,0.5$ and $0.85$ (see Figs.~\ref{fig3}(b), (d) and (h)). The lowest energy mode for each composition, denoted by open symbols, is determined by the analysis of the corresponding dispersion relation along ${\bf \Gamma}-{\bf M}-{\bf \Gamma}$ as discussed below, so that complications due to the presence of the elastic peak can be eliminated.
From Figs.~\ref{fig3}(b),(d),(f),(h), we conclude that $E_{\rm soft}^2\propto T$ for all compositions.
The proportionality factor, which determines how rapidly the relevant phonon mode softens, is not a strong function of $x$. 
The temperature at which $E_{\rm soft}^2$ extrapolates to zero is defined as $T^*$. The values of $T^*$ obtained are plotted against the nominal calcium concentration in Fig.~\ref{fig3}(i), showing an excellent agreement with the phase diagram constructed earlier using resistivity \cite{Goh2015} and specific heat \cite{Yu2015} data. In $x=0$, we additionally confirmed the recovery of the soft mode below \Tstar: the soft mode hardens again and reaches 2.4~meV at 70~K, which is $\sim$60~K below $T^*$, as shown in Supplemental Material \cite{SUPP}.
These results are fully consistent with the expectation from the Landau theory for a second-order structural transition where $E_{\rm soft}^2$ plays the role of the inverse susceptibility \cite{Dove2003}, and confirm that the transition is driven by this particular phonon mode. Intriguingly, for $x=0.85$, $E_{\rm soft}^2$ extrapolates to zero at $(-5.7 \pm 7.7)$~K, thereby establishing this composition to be at, or very close to, the SQCP. Thus, our IXS data provide the first compelling microscopic evidence for the existence of a SQCP in the phase diagram of \CaSrRhx.

For all samples studied, we additionally performed X-ray diffraction measurements at {\bf M} to probe the superlattice intensity using the same spectrometer. The temperature dependence of the superlattice intensity at {\bf M} is shown in Figs.~\ref{fig3}(a),(c),(e),(g). A rapid growth of the superlattice intensity occurs near $T^*$, where $E_{\rm soft}^2\rightarrow0$. This reinforces the view that the phonon mode softening is intimately linked to the structural transition. In $x=0.75$, the superlattice intensity begins to grow at $\sim$30~K, which is slightly higher than $T^*$ determined from the analysis of $E_{\rm soft}^2(T)$ of the soft mode. However, the superlattice intensity is weak in $x=0.75$ and the growth is rather slow. Further studies with a diffractometer (instead of a spectrometer) is highly desirable.
In $x=0.85$, we did not detect the growth of the superlattice intensity down to 7~K, the base temperature of our setup, again consistent with the measured $E_{\rm soft}^2(T)$.

\begin{figure}[!t]\centering
       \resizebox{9cm}{!}{
              \includegraphics
              {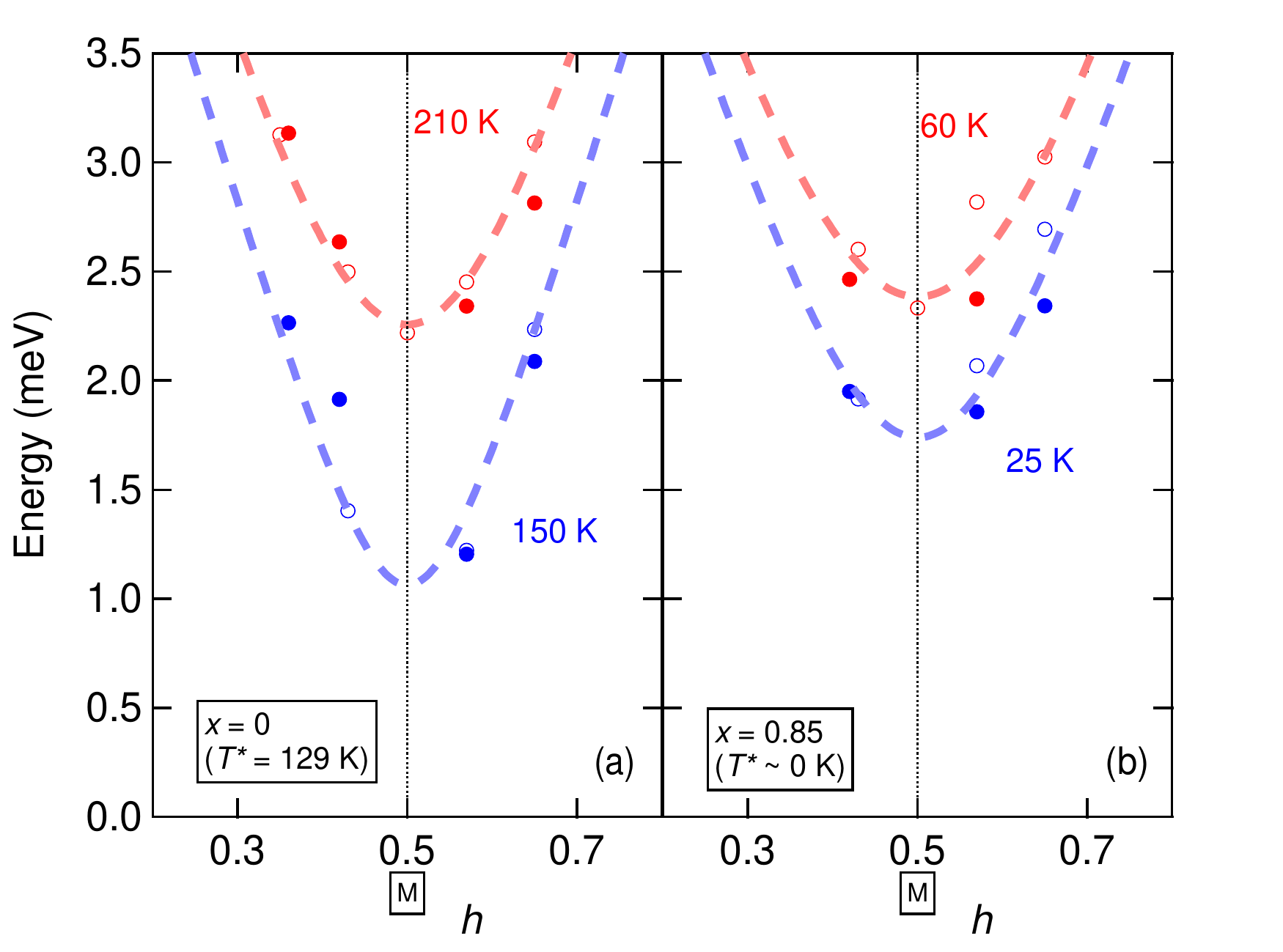}}              				
              \caption{\label{fig4} 
              The dispersion relations in (a) $x=0$, and (b) $x=0.85$ measured around $\mathbf{Q}=$ (6.5 6.5 1). Closed and open symbols represent data at $\mathbf{Q}=$ (6$+h$ 6$+h$ 1) and (6$+h$ 7$-h$ 1) respectively. The dispersion relations are analysed using Equation (1).}
\end{figure}

In order to understand how structural instabilities affect the low energy excitations, we examine the phonon modes around the {\bf M} point. As the intensity of the modes around {\bf M} is weak compared with the higher energy modes, we have performed IXS measurements at $\mathbf{Q}=$ (6$+h$ 6$+h$ 1) and (6$+h$ 7$-h$ 1), where $h=0.5$ corresponds to the {\bf M} point. Representative IXS spectra can be found in the  Supplemental Material \cite{SUPP}. IXS scans in (6$+h$ 6$+h$ 1) and (6$+h$ 7$-h$ 1) will predominantly measure modes with longitudinal and transverse component respectively. Figs.~\ref{fig4}(a),(b) show the dispersion of $x=0$ and  $x=0.85$ around {\bf M}, respectively, each at two chosen temperatures above $T^*$. Similar data for $x=0.5$ (C2) and $x=0.75$ are shown in Ref.~\cite{SUPP}. Observation of soft modes confirms the existence of lattice instability at {\bf M}. The softening of the phonon modes is broadly distributed in momentum space, in contrast to a sharp Kohn anomaly caused by Fermi surface nesting in low dimensional charge-density-wave systems \cite{Gruner2009}. The dispersion relations above \Tstar\ can be fitted by 
\begin{equation}
E(q')=\sqrt{c^2q'^2+\Delta^2},
\end{equation}
where $q'$ is the wave vector measured from {\bf M} and $\Delta$ is the energy of the soft mode at {\bf M}. As there is no significant difference between the two dispersions in both directions, data in both directions are fitted by a single curve. The fitted values of $\Delta^2$ for the lower temperature branches are shown as open symbols in Figs. \ref{fig3}(b),(d),(f),(h).

In $x=0$, which is far from the SQCP, the soft phonon mode has significantly hardened by the time superconductivity sets in \cite{SUPP}. Therefore, $2\Delta_{\rm sc}/k_\mathrm{B}T_\mathrm{c}$ and $\Delta C/\gamma T_c$ both take the BCS weak coupling values, and $T_c$ is lower. At the SQCP, the soft mode at {\bf M} only softens near the zero temperature, thereby contributing to the large $\sim T^3$ specific heat at low temperatures, strong-coupling superconductivity, and a higher $T_c$ \cite{Yu2015}.
Furthermore, the observed linear-in-$T$ resistivity \cite{Goh2015} can be naturally interpreted as the result of scattering off the low-lying phonon modes. 
To investigate the role of the $\sim$5~meV branch mentioned earlier, we also track this particular branch below $T^*$ for $x=0$ \cite{SUPP}. This branch exists below $T^*$ as well, and it is still temperature insensitive. Hence, this branch appears to be very robust across a wide temperature and composition range, and it is not relevant to the structural transition or the enhancement of superconductivity upon Ca tuning.
Thus, strong-coupling superconductivity and structural transition are driven by the same soft phonon mode in \CaSrRhx.

Recently, strong-coupling superconductivity and linear-in-$T$ resistivity were also reported in Bi-III, which has an incommensurate host-guest structure, accompanied by ``built-in''  phason modes \cite{Brown2018}. Thus, Bi-III and \CaSrRhx\ share an important similarity in that both systems feature a large phonon density-of-states at low energy that facilitates the formation of strong-coupling superconductivity, despite the different mechanisms of generating these low-lying modes. 

The IXS data reported here provide key spectroscopic evidence for the collapse of an energy scale when the second-order phase transition temperature is tuned to 0~K. Therefore, \CaSrRhx\ provides another example in which the principle of quantum criticality can guide the search of superconductivity or the optimization of superconducting properties. Note that quantum critical behavior associated with structural transition has also been studied in ferroelectric materials \cite{Rowley2014,Chandra2017} which are insulators and have net dipole moment given by the softening of a zone-center mode. 
In SrTiO$_{3-\delta}$, oxygen vacancies introduce charge carriers and induce superconductivity for small $\delta$. Through the substitution of Sr by Ca, a ferroelectric-like quantum phase transition would take place at finite $\delta$, where superconductivity is enhanced \cite{Rischau2017}. Therefore, Sr$_{1-x}$Ca$_x$TiO$_{3-\delta}$ emerges as an important paradigm to explore the interplay between the ferroelectric-like structural distortion and superconductivity. It would be interesting to follow the lattice dynamics of Sr$_{1-x}$Ca$_x$TiO$_{3-\delta}$ across the quantum phase transition, and compare the resultant phonon structure with the present study.
In addition to superconductivity, it is important to point out that other interesting phenomena have been observed when one looks closer to the zero temperature instability. For instance, near a ferromagnetic quantum phase transition, the character of the phase transition is often discontinuous, raising the hotly debated question if a ferromagetic QCP can ever be realised \cite{Brando2016}. With the notion of structural quantum criticality established in \CaSrRhx, further explorations in the vicinity of the SQCP are highly desirable.

In summary, we have probed the lattice dynamics in \CaSrRhx\ with IXS measurements. 
Far away from structural instabilities, there are low-lying phonon branches with weak dispersion at $\sim$~5--8~meV with no obvious dependence on temperature and chemical pressure. However, our data reveal another phonon branch which is more dispersive, and exhibits strong dependence on temperature and chemical pressure. In particular, in $x=0$, $0.5$, and $0.75$, we found that the soft phonon mode at {\bf M} can be completely softened close to the temperature where the superlattice intensity begins to grow, signifying its relevance to structural instabilities. For $x=0.85$, the energy squared of this phonon mode extrapolates to zero at $(-5.7 \pm 7.7)$~K, indicating that the composition is located very close to or at the SQCP. The analysis of the dispersion relation around {\bf M} reveals the existence of low-lying phonon modes, which offers a natural explanation for the low Debye temperature, the maximized $T_\mathrm{c}$ as well as the enhanced electron-phonon coupling strength near the SQCP.  Therefore, \CaSrRhx\ is a rare quantum critical system in which manifestations of phonon softening on electronic, thermal, and superconducting properties can be investigated systematically. 

\begin{acknowledgments}
The authors acknowledge Xiaoye Chen and Youichi Yanase for helpful discussions, and Hiroshi Uchiyama for experimental assistance for one of the beamtimes.
This experiment was carried out under the approval of JASRI (Proposal No. 2013B1095, 2015B1294, 2016A1160, 2017A1130, 2017B1228 and 2018A1399).
This work was financially supported by Research Grants Council of Hong Kong (ECS/24300214, GRF/14300117), CUHK Direct Grant (No. 4053223), CUHK Startup (No. 4930048), 
National Natural Science Foundation of China (No. 11504310), 
Grants-in-Aid for Scientific Research from MEXT of Japan (No. 16H04131, No. 15H03697, No. 16K05031). 
\end{acknowledgments}

\begin{thebibliography}{10}

\bibitem{Mathur1998}
N.~D. Mathur, F.~M. Grosche, S.~R. Julian, I.~R. Walker, D.~M. Freye, R.~K.~W.
  Haselwimmer, and G.~G. Lonzarich,
\newblock Nature {\bf 394}, 39 (1998).

\bibitem{Slooten2009}
E.~Slooten, T.~Naka, A.~Gasparini, Y.~K. Huang, and A.~de~Visser,
\newblock Phys. Rev. Lett. {\bf 103}, 097003 (2009).

\bibitem{Gegenwart2008}
P.~Gegenwart, Q.~Si, and F.~Steglich,
\newblock Nat. Phys. {\bf 4}, 186 (2008).

\bibitem{Paglione2010}
J.~Paglione and R.~L. Greene,
\newblock Nat. Phys. {\bf 6}, 645 (2010).

\bibitem{Shibauchi2014}
T.~Shibauchi, A.~Carrington, and Y.~Matsuda,
\newblock Annu. Rev. Condens. Matter Phys. {\bf 5}, 113 (2014).

\bibitem{Gruner2017}
T.~Gruner, D.~Jang, Z.~Huesges, R.~Cardoso-Gil, G.~H. Fecher, M.~M. Koza,
  O.~Stockert, A.~P. Mackenzie, M.~Brando, and C.~Geibel,
\newblock Nat. Phys. {\bf 13}, 967 (2017).

\bibitem{Scalapino2012}
D.~J. Scalapino,
\newblock Rev. Mod. Phys. {\bf 84}, 1383 (2012).

\bibitem{Monthoux2007}
P.~Monthoux, D.~Pines, and G.~G. Lonzarich,
\newblock Nature {\bf 450}, 1177 (2007).

\bibitem{Chu1973}
C.~W. Chu and L.~R. Testardi,
\newblock Phys. Rev. Lett. {\bf 32}, 766 (1974).

\bibitem{Chu1974}
C.~W. Chu,
\newblock Phys. Rev. Lett. {\bf 33}, 1283 (1974).

\bibitem{Testardi1975}
L.~R. Testardi,
\newblock Rev. Mod. Phys. {\bf 47}, 637 (1975).

\bibitem{Holder1992}
A.~Bussmann-Holder and A.~R. Bishop,
\newblock Zeitschrift f{\"u}r Physik B {\bf 86}, 183 (1992).

\bibitem{Birgeneau1987}
R.~J. Birgeneau, C.~Y. Chen, D.~R. Gabbe, H.~P. Jenssen, M.~A. Kastner, C.~J.
  Peters, P.~J. Picone, T.~Thio, T.~R. Thurston, H.~L. Tuller, J.~D. Axe,
  P.~B\"oni, and G.~Shirane,
\newblock Phys. Rev. Lett. {\bf 59}, 1329 (1987).

\bibitem{Wakimoto2004}
S.~Wakimoto, S.~Lee, P.~M. Gehring, R.~J. Birgeneau, and G.~Shirane,
\newblock J. Phys. Soc. Jpn. {\bf 73}, 3413 (2004).

\bibitem{Kang2011}
H.~Kang, Y.~Lee, J.~Lynn, S.~Shiryaev, and S.~Barilo,
\newblock Physica C {\bf 471}, 303  (2011).

\bibitem{Yang2012}
J.~J. Yang, Y.~J. Choi, Y.~S. Oh, A.~Hogan, Y.~Horibe, K.~Kim, B.~I. Min, and
  S.-W. Cheong,
\newblock Phys. Rev. Lett. {\bf 108}, 116402 (2012).

\bibitem{Pyon2012}
S.~Pyon, K.~Kudo, and M.~Nohara,
\newblock J. Phys. Soc. Jpn. {\bf 81}, 053701 (2012).

\bibitem{Fang2013}
A.~F. Fang, G.~Xu, T.~Dong, P.~Zheng, and N.~L. Wang,
\newblock Sci. Rep. {\bf 3}, 1153 (2013).

\bibitem{Kamitani2016}
M.~Kamitani, H.~Sakai, Y.~Tokura, and S.~Ishiwata,
\newblock Phys. Rev. B {\bf 94}, 134507 (2016).

\bibitem{Kudo2016}
K.~Kudo, H.~Ishii, and M.~Nohara,
\newblock Phys. Rev. B {\bf 93}, 140505 (2016).

\bibitem{Heil2017}
C.~Heil, S.~Ponc\'e, H.~Lambert, M.~Schlipf, E.~R. Margine, and F.~Giustino,
\newblock Phys. Rev. Lett. {\bf 119}, 087003 (2017).

\bibitem{Cruz2008}
C.~de~la Cruz, Q.~Huang, J.~W. Lynn, J.~Li, W.~R. II, J.~L. Zarestky, H.~A.
  Mook, G.~F. Chen, J.~L. Luo, N.~L. Wang, and P.~Dai,
\newblock Nature {\bf 453}, 899 (2008).

\bibitem{Yoshizawa2012}
M.~Yoshizawa, D.~Kimura, T.~Chiba, S.~Simayi, Y.~Nakanishi, K.~Kihou, C.-H.
  Lee, A.~Iyo, H.~Eisaki, M.~Nakajima, and S.~ichi Uchida,
\newblock J. Phys. Soc. Jpn. {\bf 81}, 024604 (2012).

\bibitem{Niedziela2011}
J.~L. Niedziela, D.~Parshall, K.~A. Lokshin, A.~S. Sefat, A.~Alatas, and
  T.~Egami,
\newblock Phys. Rev. B {\bf 84}, 224305 (2011).

\bibitem{Kudo2012}
K.~Kudo, M.~Takasuga, Y.~Okamoto, Z.~Hiroi, and M.~Nohara,
\newblock Phys. Rev. Lett. {\bf 109}, 097002 (2012).

\bibitem{Hirai2012}
D.~Hirai, F.~von Rohr, and R.~J. Cava,
\newblock Phys. Rev. B {\bf 86}, 100505 (2012).

\bibitem{Poudel2016}
L.~Poudel, A.~F. May, M.~R. Koehler, M.~A. McGuire, S.~Mukhopadhyay, S.~Calder,
  R.~E. Baumbach, R.~Mukherjee, D.~Sapkota, C.~de~la Cruz, D.~J. Singh,
  D.~Mandrus, and A.~D. Christianson,
\newblock Phys. Rev. Lett. {\bf 117}, 235701 (2016).

\bibitem{Handunkanda2015}
S.~U. Handunkanda, E.~B. Curry, V.~Voronov, A.~H. Said, G.~G. Guzm\'an-Verri,
  R.~T. Brierley, P.~B. Littlewood, and J.~N. Hancock,
\newblock Phys. Rev. B {\bf 92}, 134101 (2015).

\bibitem{Kase2011}
N.~Kase, H.~Hayamizu, and J.~Akimitsu,
\newblock Phys. Rev. B {\bf 83}, 184509 (2011).

\bibitem{Goh2015}
S.~K. Goh, D.~A. Tompsett, P.~J. Saines, H.~C. Chang, T.~Matsumoto, M.~Imai,
  K.~Yoshimura, and F.~M. Grosche,
\newblock Phys. Rev. Lett. {\bf 114}, 097002 (2015).

\bibitem{Yu2015}
W.~C. Yu, Y.~W. Cheung, P.~J. Saines, M.~Imai, T.~Matsumoto, C.~Michioka,
  K.~Yoshimura, and S.~K. Goh,
\newblock Phys. Rev. Lett. {\bf 115}, 207003 (2015).

\bibitem{Lue2016}
C.~S. Lue, C.~N. Kuo, C.~W. Tseng, K.~K. Wu, Y.-H. Liang, C.-H. Du, and Y.~K.
  Kuo,
\newblock Phys. Rev. B {\bf 93}, 245119 (2016).

\bibitem{Cheung2016}
Y.~W. Cheung, J.~Z. Zhang, J.~Y. Zhu, W.~C. Yu, Y.~J. Hu, D.~G. Wang, Y.~Otomo,
  K.~Iwasa, K.~Kaneko, M.~Imai, H.~Kanagawa, K.~Yoshimura, and S.~K. Goh,
\newblock Phys. Rev. B {\bf 93}, 241112 (2016).

\bibitem{Cheung2017}
Y.~W. Cheung, Y.~J. Hu, S.~K. Goh, K.~Kaneko, S.~Tsutsui, P.~W. Logg, F.~M.
  Grosche, H.~Kanagawa, Y.~Tanioku, M.~Imai, T.~Matsumoto, and K.~Yoshimura,
\newblock J. Phys.: Confer. Ser. {\bf 807}, 032002 (2017).

\bibitem{Hu2017}
Y.~J. Hu, Y.~W. Cheung, W.~C. Yu, M.~Imai, H.~Kanagawa, J.~Murakawa,
  K.~Yoshimura, and S.~K. Goh,
\newblock Phys. Rev. B {\bf 95}, 155142 (2017).

\bibitem{Klintberg2012}
L.~E. Klintberg, S.~K. Goh, P.~L. Alireza, P.~J. Saines, D.~A. Tompsett, P.~W.
  Logg, J.~Yang, B.~Chen, K.~Yoshimura, and F.~M. Grosche,
\newblock Phys. Rev. Lett. {\bf 109}, 237008 (2012).

\bibitem{Biswas2015}
P.~K. Biswas, Z.~Guguchia, R.~Khasanov, M.~Chinotti, L.~Li, K.~Wang,
  C.~Petrovic, and E.~Morenzoni,
\newblock Phys. Rev. B {\bf 92}, 195122 (2015).

\bibitem{Feng2012}
Y.~Feng, J.~Wang, R.~Jaramillo, J.~van Wezel, S.~Haravifard, G.~Srajer, Y.~Liu,
  Z.-A. Xu, P.~B. Littlewood, and T.~F. Rosenbaum,
\newblock Proc. Natl. Acad. Sci. USA {\bf 109}, 7224 (2012).

\bibitem{SUPP}
See Supplemental Material for (1) experimental methods, (2) procedures of
  analyzing the IXS spectra, (3) IXS spectra showing phonon softening for
  $x=0.5$ and $x=0.85$, (4) dispersion relations of $x=0.5$ and $x=0.75$ and (5) phonon mode recovery below $T^*$ for $x=0$.

\bibitem{Yang2010}
J.~Yang, B.~Chen, C.~Michioka, and K.~Yoshimura,
\newblock J. Phys. Soc. Jpn. {\bf 79}, 113705 (2010).

\bibitem{Tompsett2014}
D.~A. Tompsett,
\newblock Phys. Rev. B {\bf 89}, 075117 (2014).

\bibitem{Mazzone2015}
D.~G. Mazzone, S.~Gerber, J.~L. Gavilano, R.~Sibille, M.~Medarde, B.~Delley,
  M.~Ramakrishnan, M.~Neugebauer, L.~P. Regnault, D.~Chernyshov, A.~Piovano,
  T.~M. Fern\'andez-D\'{\i}az, L.~Keller, A.~Cervellino, E.~Pomjakushina,
  K.~Conder, and M.~Kenzelmann,
\newblock Phys. Rev. B {\bf 92}, 024101 (2015).

\bibitem{Dove2003}

\newblock M.~T. Dove{\em $,\ $Structure and dynamics: an atomic view of
  materials} Vol.~1 (Oxford University Press, 2003).

\bibitem{Gruner2009}
G.~Gruner,
\newblock {\em Density Waves In Solids$\ $}Frontiers in Physics (Westview
  Press, 2009).

\bibitem{Brown2018}
P.~Brown, K.~Semeniuk, D.~Wang, B.~Monserrat, C.~J. Pickard, and F.~M. Grosche,
\newblock Sci. Adv. {\bf 4}, eaao4793 (2018).

\bibitem{Rowley2014}
S.~E. Rowley, L.~J. Spalek, R.~P. Smith, M.~P.~M. Dean, M.~Itoh, J.~F. Scott,
  G.~G. Lonzarich, and S.~S. Saxena,
\newblock Nat. Phys. {\bf 10}, 367 (2014).

\bibitem{Chandra2017}
P.~Chandra, G.~G. Lonzarich, S.~E. Rowley, and J.~F. Scott,
\newblock Reports on Progress in Physics {\bf 80}, 112502 (2017).

\bibitem{Rischau2017}
C.~W. Rischau, X.~Lin, C.~P. Grams, D.~Finck, S.~Harms, J.~Engelmayer,
  T.~Lorenz, Y.~Gallais, B.~Fauqu\'e, J.~Hemberger, and K.~Behnia,
\newblock Nat. Phys. {\bf 13}, 643 (2017).

\bibitem{Brando2016}
M.~Brando, D.~Belitz, F.~M. Grosche, and T.~R. Kirkpatrick,
\newblock Rev. Mod. Phys. {\bf 88}, 025006 (2016).

\end{thebibliography}

\end{document}